\documentclass[12pt]{article}
\usepackage{amsmath}
\usepackage{bm}
\usepackage{color}
\usepackage{braket}
\usepackage{caption}
\usepackage{subcaption}
\usepackage{graphicx}
\usepackage{amssymb}
\usepackage{float}
\usepackage{qcircuit}
\begin{document}
\begin{center}
\begin{large}
{\bf Properties of multi-qubit variational quantum states representing weighted graphs and their computing with quantum programming}
\end{large}
\end{center}

\centerline {Kh. P. Gnatenko \footnote{E-Mail address: khrystyna.gnatenko@gmail.com}$^{,2}$, A. Kaczmarek \footnote{E-Mail address: akacz@softserveinc.com}} 
\medskip

\centerline {$^1$\small \it Ivan Franko National University of Lviv,}
\centerline {\small \it Professor Ivan Vakarchuk Department for Theoretical Physics,}
\centerline {\small \it 12 Drahomanov St., Lviv, 79005, Ukraine}

\centerline{$^2$\small \it SoftServe Inc.,  Austin, TX, USA.}

\abstract{ 
  We study multi-qubit variational quantum states that can be considered as vertex- and edge-weighted graph. These states are constructed as single-layer variational circuits with $RX$ rotations and $RZZ$ entangling gates, corresponding to graphs of arbitrary structure. In general case of quantum graph states of arbitrary structure we derive the geometric measure of entanglement and evaluate quantum correlators. It is shown that these quantities are related to the edge-weight structure around the corresponding vertices in the graph (i.e., edge weights incident to the vertices and vertex weights associated with their closed neighborhoods). In the special case of quantum states representing unweighted graphs, these quantities are related to the degrees of the corresponding vertices in the graph. As an example, we analyze the state associated with the star graph $K_{1,4}$ using noisy  quantum computing on the AerSimulator. The results are in good agreement with theoretical predictions. These findings demonstrate a connection between graph structure and quantum properties, enabling the study of properties of classical graphs via quantum computing.

}

\section{Introduction}

Quantum computing is based on operations on multi-qubit entangled quantum states.
Entanglement is widely recognized as a fundamental resource for advanced information processing tasks in quantum computing and quantum communications. It enables powerful protocols such as quantum cryptography and teleportation, and underpins the exponential speed-ups of many quantum algorithms \cite{1_PhysRevLett.67.661, 2_PhysRevLett.70.1895, 3_10.1098/rspa.2002.1097}. Accordingly, considerable effort has been devoted to both theoretically quantifying entanglement and generating high-quality entangled states on actual devices \cite{4_RevModPhys.81.865}. In recent years, experiments have demonstrated genuinely multi-partite entangled states on superconducting quantum processors (e.g., entangling all 16 qubits of an IBM device and observing multi-qubit entanglement in 20-qubit systems \cite{5_Wang2018, 6_PhysRevX.8.021012}). To systematically study such phenomena, researchers have proposed various measures of entanglement. In this work, we focus on the geometric measure of entanglement (GME) introduced by Shimony \cite{Shimony1995}, defined as the minimum squared Fubini–Study distance between an n-qubit state and the set of all n-qubit separable (unentangled) states. This measure has an intuitive geometric meaning and, importantly, has been linked to measurable quantities in experiments. In particular, GME can be evaluated via the expectation values of spin (Pauli) operators – a relationship was shown in \cite{Frydryszak2017}, which lays the foundation for entanglement estimation on quantum hardware \cite{Kuzmak_2020,Gnatenko_2021,Gnatenko_2024}. 

Beyond bipartite entanglement, an ongoing research challenge is to understand how entanglement is distributed in multipartite quantum states, especially those prepared by modern parameterized quantum circuits \cite{7_GUHNE20091,Vesperini2023}. A foundational result presented in \cite{McClean2018} revealed that highly expressive parameterized quantum circuits can suffer from barren plateaus—regions of vanishing gradients that hinder trainability—due to excessive entanglement. The authors of paper \cite{Sim2019} introduced quantitative descriptors for expressibility and entangling capability, showing how circuit architecture and gate choices influence the ability of PQCs to generate entangled states. In  \cite{Cerezo2021} it was further demonstrated that the emergence of barren plateaus depends critically on the cost function and circuit depth, with local cost functions mitigating the issue in shallow circuits. Complementing these theoretical insights, authors of paper \cite{Hubregtsen2021} empirically studied the relationship between entanglement, expressibility, and classification accuracy in PQCs, finding that while expressibility correlates with performance, entanglement alone is not always predictive of success.

One prominent family of highly entangled multipartite states are quantum graph states, which are generated by applying two-qubit entangling gates (such as controlled-phase gates) between selected pairs of qubits initialized in a product state. Graph states play a central role in quantum error correction, cryptography, and network protocols, and serve as a natural testbed for studying multi-qubit entanglement structure \cite{8_PhysRevLett.86.5188, 9_PhysRevA.78.042309}. Prior works have derived closed-form GME values for certain graph states, finding that a qubit’s entanglement is often determined by its connectedness in the interaction graph \cite{Gnatenko_2021, Gnatenko_2024, Gnatenko2026}.  Recent work has also proposed variational algorithms that directly optimize over entanglement witnesses or geometric entanglement bounds, enabling scalable estimation of multipartite entanglement in noisy intermediate-scale quantum (NISQ) devices \cite{AzimiMousolou2025}.

Building on  prior work, we examine single-layer variational quantum states constructed using $RX$ gates and entangling blocks composed of $RZZ$ gates applied in an arbitrary manner. We analytically calculate both the entanglement of these states and the corresponding quantum correlators. We establish the dependence of these quantities on the classical parameters associated with the vertices and edges of the corresponding graphs. In the specific case of star graphs, we compute the entanglement and quantum correlations using noisy quantum computing on the AerSimulator.

The paper is organized as follows. In Section \ref{2} we calculate analytically the geometric measure of entanglement of quantum graph states and study quantum correlators in general case of graphs of arbitrary structure. In Section \ref{3} we present results of quantum calculations of the entanglement of single-layer variational quantum states representing weighted graphs as well as and quantum computing of quantum correlators. Conclusions are presented in Section 4.

\section{Properties of one-layer variational quantum states corresponding to weighted graphs}\label{2}
\begin{figure}[H]
    \centering
    \includegraphics[scale=0.55]{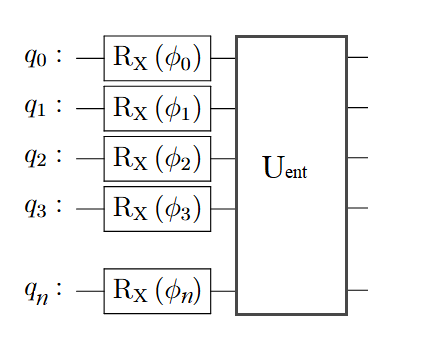}
    \caption{Single-layer variational quantum protocol with rotational block represented with $RX$ gates and entangled block $U_{ent}=\prod_{j,k}RZZ_{jk}$.}
    \label{prot}
\end{figure}

Let us  study single-layer variational quantum states constructed using $RX_j(\phi_i)=\exp(-i\phi_j\sigma^x_j/2)$ gates, and entangled block $U_{ent}$ represented  with $RZZ_{jk}(\theta_{jk})=\exp(-i\theta_{jk}\sigma^z_j\sigma^z_k/2)$ gates applied with arbitrary topology, $U_{ent}=\prod_{j,k}RZZ_{jk}$ (see Fig.  \ref{prot}). 

The resulting states can be interpreted as quantum graph states. They can be represented by vertex- and edge-weighted graph
$G(V,E)$, where the vertices 
$V$
correspond to qubits and the action of two-qubit gates is represented by edges 
$E$, the weights of edges $(j,k)$ are characterized by parameters $\theta_{jk}$. The weights of vertexes are characterized with parameters $\phi_j$.  We consider a single-layer variational quantum state that corresponds to a quantum graph state representing an arbitrary graph (with 
$RZZ$ gates applied in an arbitrary manner). The state is given by:

\begin{equation}
\ket{\psi_G}=\prod_{(j,k)\in E}RZZ_{jk}(\theta_{jk})\prod_{i\in V}RX_i(\phi_i)\ket{00...0}.
\end{equation}

 Let us study properties of the state.
 First, we quantify the GME of one qubit $q[l]$  with others qubits in state $\ket{\psi_I}$.  It reads 
\begin{eqnarray}
E_l (\ket{\psi_G}) = 
\frac{1}{2}\left(1-\sqrt{\langle\sigma^x_l\rangle^2+\langle\sigma^y_l\rangle^2+\langle\sigma^z_l\rangle^2}\right), \label{ed}
\end{eqnarray}
with $\langle\sigma^{\alpha}_{l}\rangle=\bra{\psi_G}\sigma^{\alpha}_{l}\ket{\psi_G}^2$ \cite{Shimony1995, Frydryszak2017}.
For the mean values of the Pauli operators, we obtain
\begin{equation}
\langle \sigma^x_{l} \rangle = \bra{00...0}\prod_{r\in V}RX^+_r(\phi_r)\prod_{j \in N_{G}(l)}RZZ^+_{jl}(2\theta_{jl})\prod_{p\in V}RX_p(\phi_p) \sigma^x_{l} \ket{00...0}=\nonumber\\
\end{equation}
\begin{equation}
=\sin \phi_l\Im\bigg[\prod_{j \in N_{G}(l)} \bigg(\cos \theta_{lj} + i\sin \theta_{lj}\cos \phi_j\bigg)\bigg].
\label{eq:sx}
\end{equation}
Here $N_{G}(l)$ is the neighborhood of vertex $l$ in graph $G$ (set of all vertices in $G$ that are connected to $l$ by an edge). Similarly, for $\langle \sigma^y_{l} \rangle$ we find
\begin{equation}
\langle \sigma^y_{l} \rangle =  \bra{00...0}\prod_{r\in V}RX^+_r(\phi_r)\prod_{j \in N_{G}(l)}RZZ^+_{jl}(2\theta_{jl})\sigma^y_{l} \prod_{p\in V} RX_p(\phi_p) \ket{00...0}=\nonumber\\
\end{equation}
\begin{equation}
=-\sin \phi_l\Re\bigg[\prod_{j \in N_{G}(l)} \bigg(\cos \theta_{lj} + i\sin\theta_{lj}\cos \phi_j\bigg)\bigg].
\label{eq:sy}
\end{equation}
And for $\langle \sigma^{z}_{l} \rangle$ we obtain
\begin{equation}\langle \sigma^{z}_{l} \rangle = \bra{0}RX^+_l(\phi_l)\sigma^z_{l}  RX_l(\phi_l) \ket{0}=\cos\phi_{l}.   
\label{eq:sz}
\end{equation}

For the entanglement of qubit with other qubits in state $\psi_{I}$ we have the following expression

\begin{eqnarray}
E_l (\ket{\psi_G})=\frac{1}{2}-\frac{1}{2}\sqrt{\cos^2\phi_{l}+\sin^2\phi_{l}\prod_{j \in N_{G}(l)}\bigg(\cos^2 \theta_{lj} + \sin^2\theta_{lj}\cos^2 \phi_j\bigg)}.\label{eg}
\end{eqnarray}
Based on the obtained expression for the entanglement,  we can conclude that these quantum property is related with  the weights of the edges connected to the vertex 
 $l$, describing by  parameters $\theta_{lk}$, $k\in N(l)$
 and also weights of vertices that correspond to the close neighborhood of $l$  $N[l]$ (all vertices adjacent to $l$ and and the vertex $l$,  itself) described with parameters $\phi_j$, $j\in N[l]$.
 
In particular case, that is $\phi_j =  \phi$, and $\theta_{jl} =  \theta$, the results for the entanglement simplifies to the following form
\begin{eqnarray}
E_l (\ket{\psi_G})=\frac{1}{2}-\frac{1}{2}\sqrt{\cos^2\phi+\sin^2\phi\bigg(\cos^2 \theta + \sin^2\theta\cos^2 \phi\bigg)^{|N_{G}(l)|}}.\label{e}
\end{eqnarray}
So, the entanglement of qubit $q[l]$ with other qubits in quantum state $\ket{\psi_G}$ depends on  the  vertex degree $|N_{G}(l)|$ (number of edges connected to vertex $l$).

Let us also calculate the correlators $\bra{\psi_G}\sigma^{\alpha}_{l} \sigma^{\beta}_{m} \ket{\psi_G}=\langle \sigma^{\alpha}_{l} \sigma^{\beta}_{m} \rangle$, $\alpha, \beta =(x,y,z)$ in general case of quantum graph state representing graph of arbitrary structure.
 For $\langle \sigma^{x}_{l} \sigma^{z}_{m} \rangle$ we have

\begin{equation}
\langle \sigma^{x}_{l} \sigma^{z}_{m} \rangle = \bra{00...0}\prod_{r\in V}RX^+_r(\phi_r)\prod_{j \in N_{G}(l)}RZZ^+_{jl}(2\theta_{jl})\sigma^x_{l} \sigma^{z}_{m} \prod_{p\in V} RX_p(\phi_p) \ket{00...0}=\nonumber\\
\end{equation}

\begin{equation}
\sin \phi_l\cos \theta_{lm}\cos \phi_m\Im\bigg[\prod_{j \in N_{G}(l)\setminus \{m\}} \bigg(\cos \theta_{lj} + i\sin \theta_{lj}\cos \phi_j\bigg)\bigg]+ \nonumber\\ \label{cg1}
\end{equation}
\begin{equation}
 + \sin\phi_l\sin\theta_{lm}\Re\bigg[\prod_{j \in N_{G}(l)\setminus \{m\}} \bigg(\cos\theta_{lj} + i\sin \theta_{lj}\cos\phi_j\bigg)\bigg].  
\end{equation}
Here $N_{G}(l)\setminus \{m\}$ denotes 
set of all vertices in $G$ except $m$ that are connected to $l$ by an edge.
Also, for  $\langle \sigma^{x}_{l} \sigma^{y}_{m}  \rangle $  we can write
\begin{eqnarray}
\langle \sigma^{x}_{l} \sigma^{y}_{m}  \rangle =
\bra{00...0}\prod_{r\in V}RX^+_r(\phi_r)\prod_{j \in N_{G}(l)\setminus{m}}  RZZ^+_{jl}(2\theta_{jl})  \prod_{k \in N_{G}(m)\setminus{l}}RZZ^+_{mk}(2\theta_{mk})\times \nonumber\\ \times \sigma^x_{l} \sigma^{y}_{m}  \prod_{p\in V} RX_p(\phi_p) \ket{00...0},\nonumber\\
\end{eqnarray}
Upon calculating the relevant expectation values, we find
\begin{eqnarray}
\langle \sigma^{x}_{l} \sigma^{y}_{m}  \rangle=\frac{1}{2}\sin\phi_l\sin\phi_m \bigg(\Im\bigg[\bigg(\cos(\theta_{lr}-\theta_{mr}) + i\sin(\theta_{lr}-\theta_{mr})\cos\phi_r\bigg) \times \nonumber \\
\times \prod_{j \in N_{G}(l)\setminus \{m\}} \prod_{k \in N_{G}(m)\setminus \{l\}, k\neq j} \bigg(\cos\theta_{lj} + i\sin\theta_{lj}\cos\phi_j\bigg) \bigg(\cos\theta_{mk} - i\sin\theta_{mk}\cos\phi_k\bigg)+\nonumber\\
 + \bigg(\cos(\theta_{lr}+\theta_{mr}) + i\sin(\theta_{lr}+\theta_{mr})\cos\phi_r\bigg) \times \nonumber \\ 
\times \prod_{j \in N_{G}(l)\setminus \{m\}} \prod_{k \in N_{G}(m)\setminus \{l\}, k\neq j} \bigg(\cos\theta_{lj} + i\sin\theta_{lj}\cos\phi_j\bigg) \bigg(\cos\theta_{mk} + i\sin\theta_{mk}\cos\phi_k\bigg)\bigg]\bigg).\nonumber\\
\end{eqnarray}
Here $r$ is the intersection of the neighborhoods of vertices $l$, $m$ in graph $G(E,V)$.
Similarly, for other correlators $\langle \sigma^{\alpha}_{l} \sigma^{\beta}_{m}  \rangle $, $\alpha, \beta =(x,y,z)$ in general case of quantum graph state representing graph of arbitrary structure we find the following results

\begin{eqnarray}
\langle \sigma^{x}_{l} \sigma^{x}_{m}  \rangle = \frac{1}{2}\sin\phi_l\sin\phi_m \bigg(\Re\bigg[\bigg(\cos(\theta_{lr}-\theta_{mr}) + i\sin(\theta_{lr}-\theta_{mr})\cos\phi_r\bigg) \times \nonumber \\
 \prod_{j \in N_{G}(l)\setminus \{m\}} \prod_{k \in N_{G}(m)\setminus \{l\}, k\neq j} \bigg(\cos\theta_{lj} + i\sin\theta_{lj}\cos\phi_j\bigg) \bigg(\cos\theta_{mk} - i\sin\theta_{mk}\cos\phi_k\bigg)-\nonumber\\
- \bigg(\cos(\theta_{lr}+\theta_{mr}) + i\sin(\theta_{lr}+\theta_{mr})\cos\phi_r\bigg) \times \nonumber\\
\prod_{j \in N_{G}(l)\setminus \{m\}} \prod_{k \in N_{G}(m)\setminus \{l\}, k\neq j} \bigg(\cos\theta_{lj} + i\sin\theta_{lj}\cos\phi_j\bigg) \bigg(\cos\theta_{mk} + i\sin\theta_{mk}\cos\phi_k\bigg)\bigg]\bigg),\nonumber \\  
\end{eqnarray}

\begin{eqnarray}
\langle \sigma^{y}_{l} \sigma^{z}_{m}  \rangle = \nonumber \\
 = -\sin \phi_l\cos \theta_{lm}\cos\phi_m\Re\bigg[\prod_{j \in N_{G}(l)\setminus \{m\}} \bigg(\cos \theta_{lj}  + i\sin \theta_{lj}\cos\phi_j\bigg)\bigg] + \nonumber \\
 \quad \quad + \sin \phi_l \sin \theta_{lm}\Im\bigg[\prod_{j \in N_{G}(l)\setminus \{m\}} \bigg(\cos \theta_{mj}  + i\sin \theta_{mj} \cos \phi_j \bigg)\bigg],   
\end{eqnarray}

\begin{eqnarray}
\langle \sigma^{y}_{l} \sigma^{y}_{m}  \rangle = \frac{1}{2}\sin\phi_l\sin\phi_m \bigg(\Re\bigg[\bigg(\cos(\theta_{lr}-\theta_{mr}) + i\sin(\theta_{lr}-\theta_{mr})\cos\phi_r\bigg) \times \nonumber \\
\prod_{j \in N_{G}(l)\setminus \{m\}} \prod_{k \in N_{G}(m)\setminus \{l\}, k\neq j} \bigg(\cos\theta_{lj} + i\sin\theta_{lj}\cos\phi_j\bigg) \bigg(\cos\theta_{mk} - i\sin\theta_{mk}\cos\phi_k\bigg)+\nonumber\\
 + \bigg(\cos(\theta_{lr}+\theta_{mr}) + i\sin(\theta_{lr}+\theta_{mr})\cos\phi_r\bigg) \times \nonumber \\
 \prod_{j \in N_{G}(l)\setminus \{m\}} \prod_{k \in N_{G}(m)\setminus \{l\}, k\neq j} \bigg(\cos\theta_{lj} + i\sin\theta_{lj}\cos\phi_j\bigg) \bigg(\cos\theta_{mk} + i\sin\theta_{mk}\cos\phi_k\bigg)\bigg]\bigg), \nonumber \\
\end{eqnarray}

\begin{equation}
\begin{aligned}
&\langle \sigma^{z}_{l} \sigma^{z}_{m}  \rangle = \cos \phi_l\cos\phi_m.\label{cg2}
\end{aligned}
\end{equation}

Based on the obtained results for the quantum correlators $\langle \sigma^{\alpha}_l \sigma^{\beta}_m\rangle$, we find that these values depend on the weights of the edges connected to vertices 
 $l$, $m$  which are characterized by the parameters $\theta_{li}$, $\theta_{mj}$, $i\in N(l)$, $j\in N(m)$, and also weights of vertices that correspond to the close neighborhoods of $l$ and $m$ described with parameters $\phi_j$, $j\in N[l]\cup N[m]$.

In case when all of the parameters were equal for all the qubits, that is $\phi_i = \phi$ and $\theta_{ij} = \theta$  and the intersection of the neighborhoods of vertices $l$, $m$ is empty the quantum correlators are related with the degrees of vertices $l$ and $m$, representing corresponding qubits $q[l]$, $q[m]$
in a graph, $|N_G(l)|$, $|N_G(m)|$. The expressions read

\begin{eqnarray}
&\langle \sigma^{x}_{l} \sigma^{z}_{m} \rangle = \sin \phi\cos\phi\cos\theta\Im\bigg[ \bigg(\cos \theta + i\sin\theta\cos\phi\bigg)^{|N_{G}(l)|}\bigg]+ \nonumber\\
& \quad \quad + \sin\phi\sin\theta\Re\bigg[\bigg(\cos\theta + i\sin\theta\cos\phi\bigg)^{|N_{G}(l)|}\bigg], \label{c1}  
\end{eqnarray}

\begin{equation}
\begin{aligned}
&\langle \sigma^{x}_{l} \sigma^{y}_{m}  \rangle = \\
& = -\sin^2\phi\Im\bigg[\bigg(\cos\theta + i\sin\theta\cos\phi\bigg)^{|N_{G}(l)|}\bigg]\Re\bigg[\bigg(\cos\theta + i\sin\theta\cos\phi\bigg)^{|N_{G}(m)|}\bigg],
\end{aligned}
\end{equation}

\begin{equation}
\begin{aligned}
&\langle \sigma^{x}_{l} \sigma^{x}_{m}  \rangle = \\
& = \sin^2\phi\Im\bigg[\bigg(\cos\theta + i\sin \theta\cos \phi\bigg)^{|N_{G}(l)|}\bigg] \Im \bigg[ \bigg(\cos \theta + i\sin \theta \cos\phi\bigg)^{|N_{G}(m)|}\bigg],
\end{aligned}    
\end{equation}

\begin{equation}
\begin{aligned}
&\langle \sigma^{y}_{l} \sigma^{z}_{m}  \rangle = \\
& = -\sin\phi\cos \theta\cos\phi\Re\bigg[\bigg(\cos \theta  + i\sin\theta\cos \phi\bigg)^{|N_{G}(l)|}\bigg] + \\
& \quad \quad + \sin \phi \sin\theta\Im\bigg[\bigg(\cos \theta + i\sin \theta\cos\phi\bigg)^{|N_{G}(l)|}\bigg],
\end{aligned}    
\end{equation}

\begin{equation}
\begin{aligned}
&\langle \sigma^{y}_{l} \sigma^{y}_{m}  \rangle = \\
& = \sin^2 \phi\Re\bigg[\bigg(\cos\theta + i\sin \theta\cos \phi\bigg)^{|N_{G}(l)|}\bigg] \Re \bigg[ \bigg(\cos \theta + i\sin \theta \cos\phi\bigg)^{|N_{G}(m)|}\bigg].
\end{aligned} 
\label{c2}
\end{equation}

So, quantifying the quantum correlators in quantum graph states one can detect these properties of graphs with quantum programming. 

In the next Section, we present the results of our study of entanglement and quantum correlators in variational quantum graph states represented by weighted graphs, using quantum computing.

\section{Quantifying entanglement and quantum correlators in quantum graph states using quantum computing}\label{3}

Let us study  properties of the quantum graph states with quantum computing. According to the definition of the geometric measure of entanglement, to quantify the entanglement of qubit $q[l]$ with other qubits in the state $\ket{\psi_G}$, one needs to prepare the state and measure the mean values of the Pauli operators corresponding to  qubit $q[l]$ in that state.
The mean value of $\sigma^z_l$ can be found based on the results of the measurement of the state of qubit $q[l]$ in the standard basis.  To find $\langle\sigma^x_l \rangle$, $\langle\sigma^y_l \rangle$, taking into account identities $\sigma^x_k=\exp(-i\pi\sigma^y/4)\sigma^z_k\exp(i\pi\sigma^y/4)$, $\sigma^y_k=\exp(i\pi\sigma^x/4)\sigma^z_k\exp(-i\pi\sigma^x/4)$ we have to apply the operator of rotation before the measurement in the standard basis. The mean values read
$\langle\sigma^x_k \rangle=\vert \langle \tilde\psi^{y} \vert 0 \rangle \vert^2-\vert \langle \tilde\psi^{y} \vert 1 \rangle \vert^2$, $\langle\sigma^y_k \rangle=\vert \langle \tilde\psi^{x} \vert 0 \rangle \vert^2-\vert \langle \tilde\psi^{x} \vert 1 \rangle \vert^2$,
with  $\vert\tilde\psi^{y}\rangle=RY_k(-\pi/2)\vert\psi_G\rangle$, $\vert\tilde\psi^{x}\rangle=RX_k(\pi/2)\vert\psi_G\rangle$.

We calculate the geometric measure of entanglement for qubit $l$ that is in a center of star topology graph $G$. The structure of the quantum circuit that encodes the example $K_{1,4}$ presented at Fig. 2.

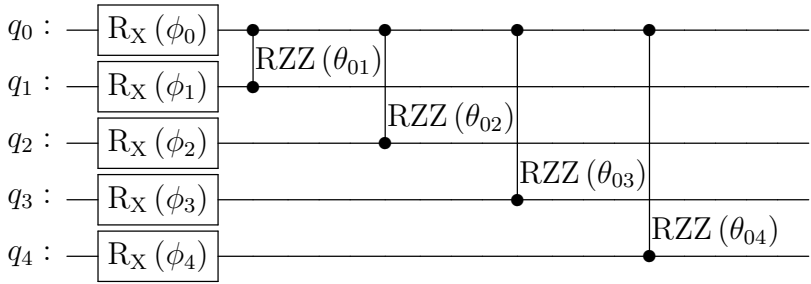
\begin{figure}[H]
\centering
\Qcircuit @C=1.0em @R=0.2em @!R { \\
	 	\nghost{{q}_{0} :  } & \lstick{{q}_{0} :  } & \gate{\mathrm{R_X}\,(\mathrm{\phi_{0}})} & \ctrl{1} & \dstick{\hspace{2.0em}\mathrm{RZZ}\,(\mathrm{\theta_{01}})} \qw & \qw & \qw & \ctrl{2} & \qw & \qw & \qw & \ctrl{3} & \qw & \qw & \qw & \ctrl{4} & \qw & \qw & \qw & \qw & \qw\\
	 	\nghost{{q}_{1} :  } & \lstick{{q}_{1} :  } & \gate{\mathrm{R_X}\,(\mathrm{\phi_{1}})} & \control \qw & \qw & \qw & \qw & \qw & \dstick{\hspace{2.0em}\mathrm{RZZ}\,(\mathrm{\theta_{02}})} \qw & \qw & \qw & \qw & \qw & \qw & \qw & \qw & \qw & \qw & \qw & \qw & \qw\\
	 	\nghost{{q}_{2} :  } & \lstick{{q}_{2} :  } & \gate{\mathrm{R_X}\,(\mathrm{\phi_{2}})} & \qw & \qw & \qw & \qw & \control \qw & \qw & \qw & \qw & \qw & \dstick{\hspace{2.0em}\mathrm{RZZ}\,(\mathrm{\theta_{03}})} \qw & \qw & \qw & \qw & \qw & \qw & \qw & \qw & \qw\\
	 	\nghost{{q}_{3} :  } & \lstick{{q}_{3} :  } & \gate{\mathrm{R_X}\,(\mathrm{\phi_{3}})} & \qw & \qw & \qw & \qw & \qw & \qw & \qw & \qw & \control \qw & \qw & \qw & \qw & \qw & \dstick{\hspace{2.0em}\mathrm{RZZ}\,(\mathrm{\theta_{04}})} \qw & \qw & \qw & \qw & \qw\\
	 	\nghost{{q}_{4} :  } & \lstick{{q}_{4} :  } & \gate{\mathrm{R_X}\,(\mathrm{\phi_{4}})} & \qw & \qw & \qw & \qw & \qw & \qw & \qw & \qw & \qw & \qw & \qw & \qw & \control \qw & \qw & \qw & \qw & \qw & \qw\\
\\ }
\label{fig:circ_alternating}
\caption{Quantum circuit for preparation of quantum state $\ket{\psi_{K_{1,4}}}$.}
\end{figure}
The corresponding quantum graph state reads
\begin{equation}
\ket{\psi_{K_{1,4}}}=\prod^4_{k=1}RZZ_{0k}(\theta_{0k})\prod^4_{i=0}RX_i(\phi_i)\ket{00000}.
\end{equation} 
Theoretical results  for the spin components required to calculate the entanglement  of qubit $q[0]$ with other qubits in quantum state $\ket{\psi_{K_{1,4}}}$ are as follows

\begin{eqnarray}
E_0(\ket{\psi_{K_{1,4}}})=\frac{1}{2}-\frac{1}{2}\sqrt{\cos^2\phi_{0}+\sin^2\phi_{0}\prod^4_{j=1}\bigg(\cos^2 \theta_{0j} + \sin^2\theta_{0j}\cos^2 \phi_j\bigg)}.
\end{eqnarray}

 With a state-of-the-art quantum simulator, we prepare the star-graph states for various circuit parameters and compute their entanglement distance via the mean-spin method, obtaining agreement with the derived formulas. We then emulate the state preparation on a noisy superconducting quantum device (incorporating realistic two-qubit gate errors and decoherence) to assess the robustness of the entanglement. For noisy simulation we transpiled the circuit to heavy hex lattice with basis gates set: $Id, X, SX, RZ, CNOT$ and noise model including  readout error probability of $10^{-2}$, X and SX gate error of order $10^{-4}$ and CNOT gate error of order $10^{-2}$. We find that the values of the entanglement are reduced under noise – as expected – but remain in qualitative agreement with the ideal predictions. This demonstrates that the approach for evaluating entanglement with mean spin is feasible on near-term quantum processors.

 We have also calculated with quantum computing on AerSimulator quantum correlators 
 $\langle \sigma_0^{\alpha} \sigma_1^{\beta}\rangle$, $\alpha,\beta =(x,y,z)$ in quantum state $\ket{\psi_{K_{1,4}}}$. Similarly, we considered noise model including  readout error probability of $10^{-2}$, X and SX gate error of order $10^{-4}$ and CNOT gate error of order $10^{-2}$. The results of quantum computing as well as absolute differences between the analytical results and results of quantum computing are presented in Figs. \ref{fig:xx_corelator}-\ref{fig:dd}. The results of quantum computing are in good agreement with the theoretical ones.

\begin{figure}[H]
    \centering
    \includegraphics[scale=0.4]{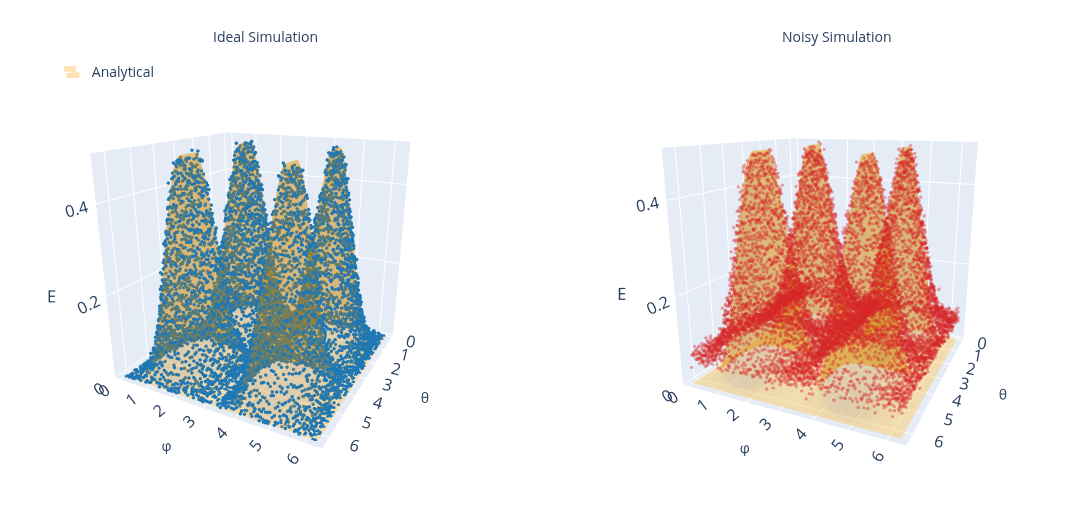}
    \caption{Surface plots of geometric measure of entanglement of qubit $q[0]$ with other qubits in quantum graph state $\ket{\psi_{K_{1,4}}}$, $\phi_i = \phi$ and $\theta_{ij} =\theta$. Surface shows analytical result, blue dots were obtained from ideal simulation on AerSimulator and red dots from noisy simulation on AerSimulator with readout error probability of $10^{-2}$, $X$ and $SX$ gate error of order $10^{-4}$ and CNOT gate error of order $10^{-2}$.}
    \label{fig:E_surface}
\end{figure}

\begin{figure}[H]
    \centering
    \includegraphics[scale=0.5]{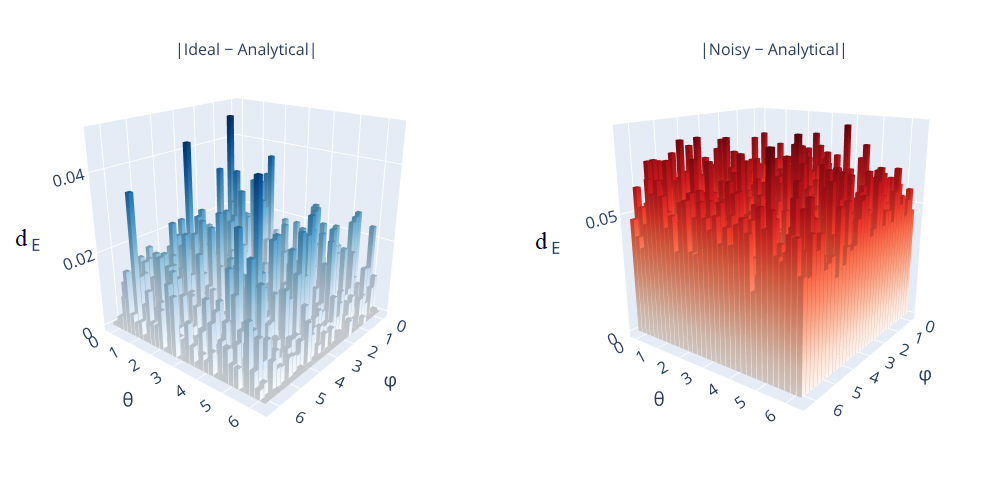}
    \caption{Bar plots of absolute differences between the analytical results for geometric measure of entanglement of qubit $q[0]$ with other qubits in quantum graph state $\ket{\psi_{K_{1,4}}}$, $\phi_i = \phi$ and $\theta_{ij} =\theta$ and results of quantum computing on AerSimulator in the case of ideal noisy simulations, $d_{E}=|E_{theor}-E_{comput}|$.}
    \label{fig:dd}
\end{figure}

\begin{figure}[H]
    \centering
    \includegraphics[scale=0.65]{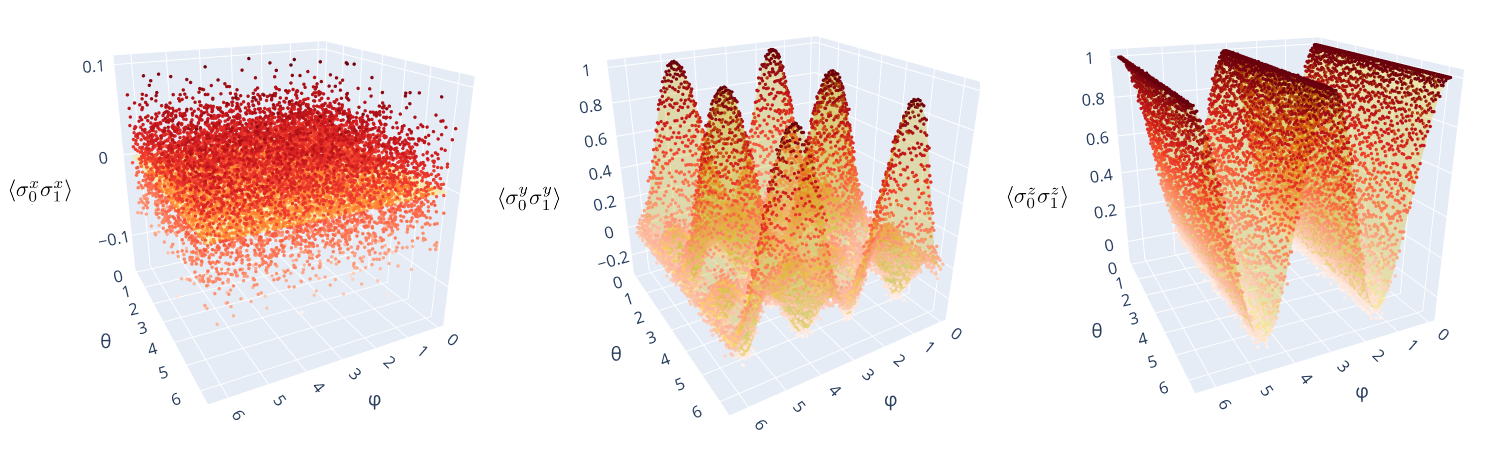}
    \caption{Surface plots of $\langle \sigma_0^{x} \sigma_1^{x}  \rangle$, $\langle \sigma_0^{y} \sigma_1^{y}  \rangle$, $\langle \sigma_0^{z} \sigma_1^{z}  \rangle$  in quantum state $\ket{\psi_{K_{1,4}}}$. Surface shows analytical result, and  dots are obtained from noisy quantum computing on AerSimulator.}
    \label{fig:xx_corelator}
\end{figure}

\begin{figure}[H]
    \centering
    \includegraphics[scale=0.42]{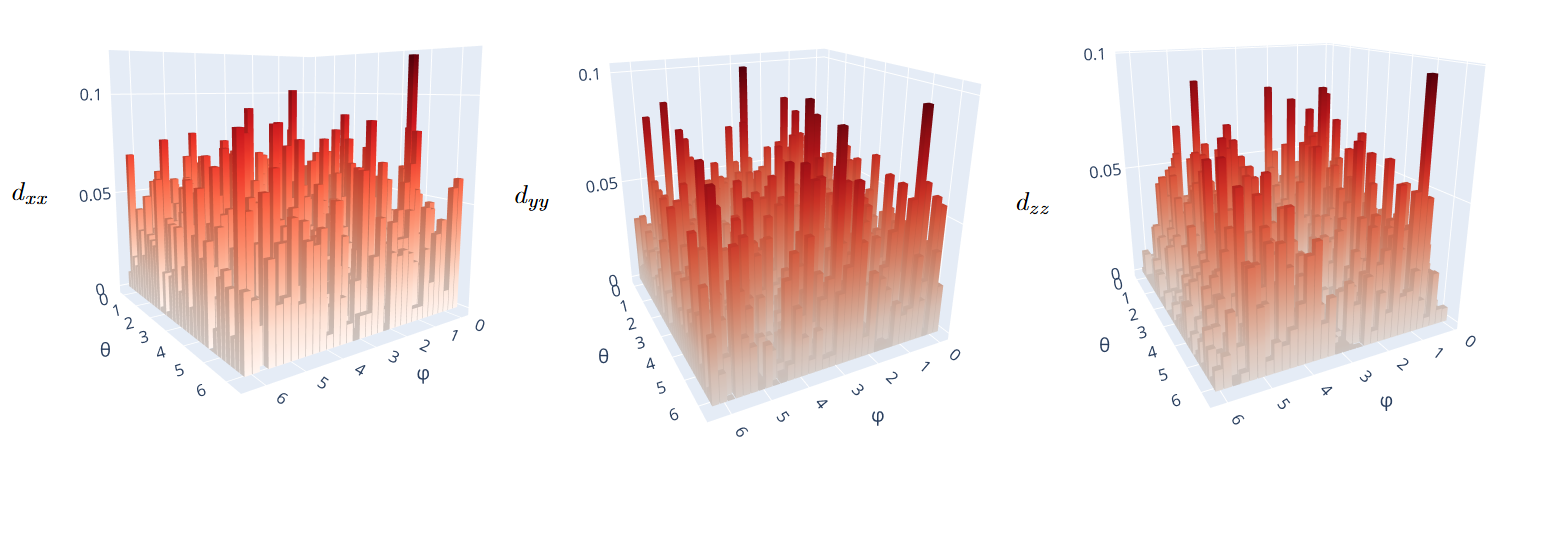}
    \caption{Bar plots of absolute differences between the theoretical results for $\langle \sigma_0^{x} \sigma_1^{x}  \rangle$, $\langle \sigma_0^{y} \sigma_1^{y}  \rangle$, $\langle \sigma_0^{z} \sigma_1^{z}  \rangle$ in quantum state $\ket{\psi_{K_{1,4}}}$ and results of noisy quantum computing on AerSimulator, $d_{\alpha \alpha}=|\langle \sigma_0^{\alpha} \sigma_1^{\alpha}\rangle_{theor}-\langle \sigma_0^{\alpha} \sigma_1^{\alpha}  \rangle_{comput}|$, $\alpha=(x,y,z)$.}
\end{figure}

\begin{figure}[H]
    \centering
    \includegraphics[scale=0.65]{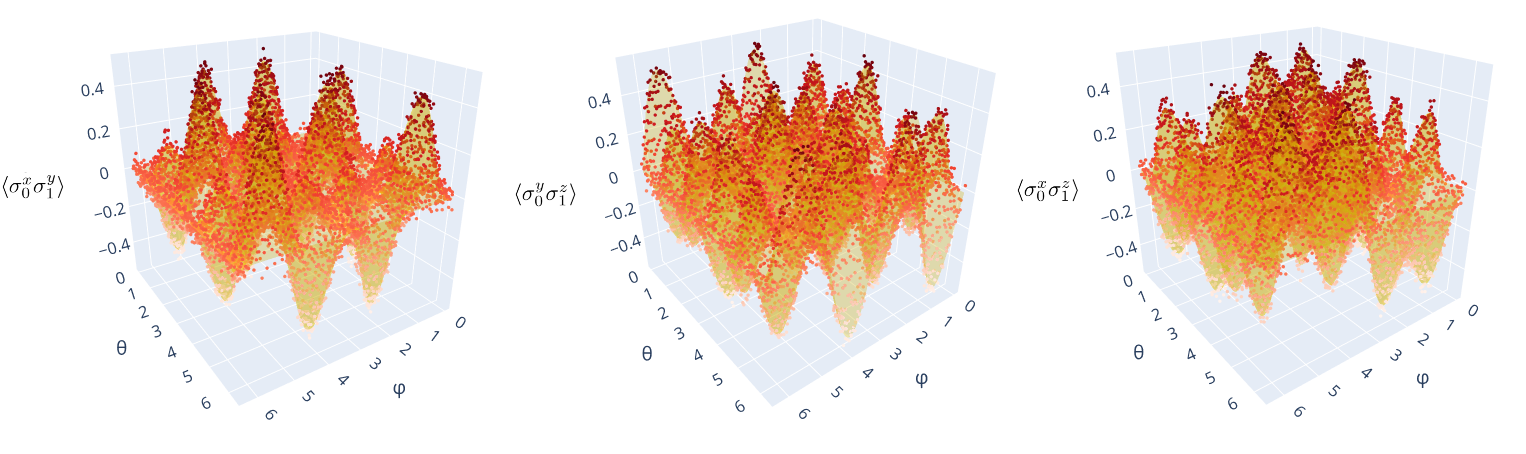}
    \caption{Surface plots of $\langle \sigma_0^{x} \sigma_1^{y}  \rangle$, $\langle \sigma_0^{y} \sigma_1^{z}  \rangle$, $\langle \sigma_0^{x} \sigma_1^{z}  \rangle$  in quantum state $\ket{\psi_{K_{1,4}}}$. Surface shows analytical result, and  dots are obtained from noisy quantum computing on AerSimulator.}
\end{figure}

\begin{figure}[H]
    \centering
    \includegraphics[scale=0.42]{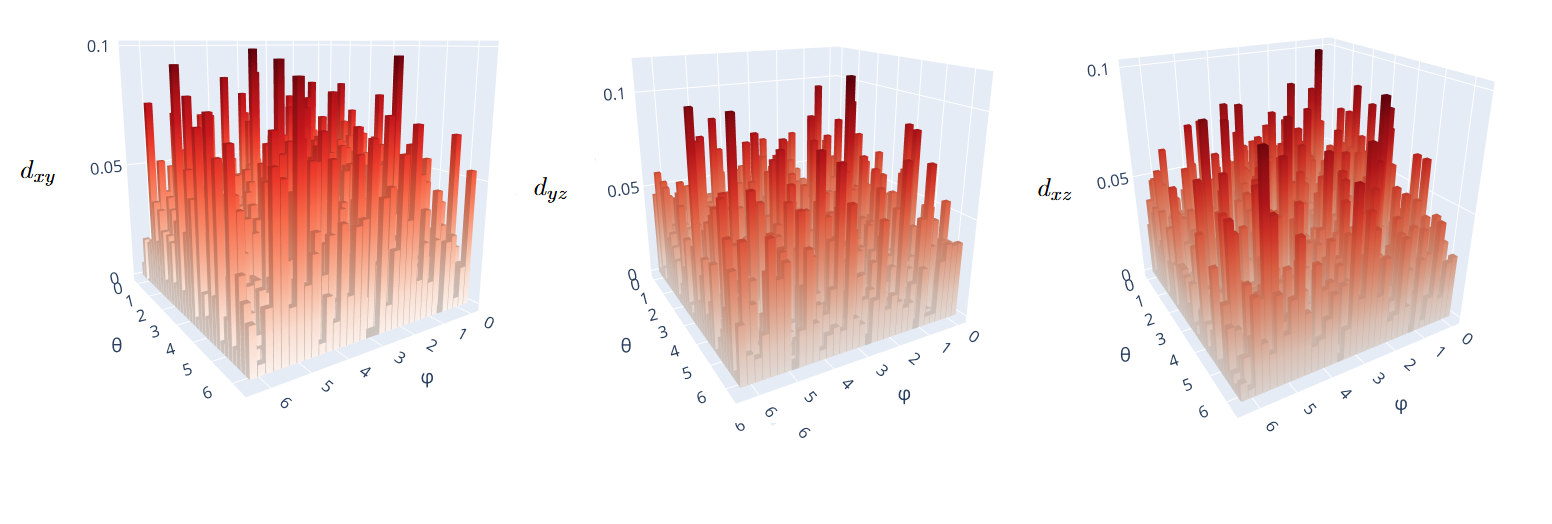}
    \caption{Bar plots of absolute differences between the analytical results for $\langle \sigma_0^{x} \sigma_1^{y}  \rangle$, $\langle \sigma_0^{y} \sigma_1^{z}  \rangle$, $\langle \sigma_0^{x} \sigma_1^{z}  \rangle$  in quantum state $\ket{\psi_{K_{1,4}}}$ and results of noisy quantum computing on AerSimulator, $d_{\alpha \beta}=|\langle \sigma_0^{\alpha} \sigma_1^{\beta}\rangle_{theor}-\langle \sigma_0^{\alpha} \sigma_1^{\beta}  \rangle_{comput}|$, $\alpha, \beta=(x,y,z)$.}
    \label{fig:dd}
\end{figure}

\section{Conclusions} 

Multi-qubit quantum states that can be considered as weighted quantum graph states have been examined. The states are variational quantum states constructed with rotational blocks using $RX$ gates and entangled blocks with $RZZ$ gates applied in an arbitrary topology. These states correspond to weighted graphs of arbitrary structure, in which vertices are represented by qubits, edges are represented by the action of $RZZ$ gates, and the weights of the edges are characterized by the parameters of the $RZZ$ gates.

For these multi-qubit quantum graph states representing graphs of arbitrary structure, we have obtained the geometric measure of entanglement as well as quantum correlators $\langle \sigma_l^{\alpha} \sigma_m^{\beta}\rangle$, where $\alpha,\beta =(x,y,z)$. It was found that these quantum quantities depend on the classical properties of the vertices and edges in the corresponding graphs.  Based on the obtained expression for the entanglement of qubit $q[l]$ with other qubits in quantum graph state, we conclude that this quantum property is governed by the edge weights incident to the vertex $l$, and the vertex weights associated with the closed neighborhood of $l$, $N[l]$ (\ref{eg}).
From the obtained results for the quantum correlators $\langle \sigma^{\alpha}_l \sigma^{\beta}_m\rangle$, we conclude that these quantities are  determined by the edge-weight structure around the vertices $l$ and $m$. In particular, they  depend on weights of edges incident to $l$ and $m$ and  vertex-weights contributions arising from the closed neighborhoods of $l$ and $m$ (\ref{cg1})-(\ref{cg2}). In particular case of unweighted graph,
the entanglement of qubit $q[l]$ in the quantum graph state is related to the degree of the corresponding vertex in the graph, while the correlators $\langle \sigma_l^{\alpha} \sigma_m^{\beta}\rangle$ depend on the degrees of vertices $l$ and $m$ (\ref{e}), (\ref{c1})-(\ref{c2}).

The obtained results are fundamental and can be used for further studies of problems of quantum information, because the properties of multi-qubit states are important resources for quantum computing. Moreover, these results open up the possibility of studying properties of classical objects such as graphs using quantum computing.

In the particular case of a quantum state corresponding to the graph $K_{1,4}$, we have studied the entanglement as well as quantum correlators using quantum computing on the AerSimulator (see Figs. 3-8). The effect of noise on the results of quantum calculations was also examined. The results of quantum simulations are in good agreement with the theoretical predictions.

\bibliographystyle{ieeetr}
	\bibliography{bibliography}
\end{document}